\pdfoutput=1
\documentclass[letterpaper, 10 pt, conference]{ieeeconf}  % Comment this line out if you need a4paper

\IEEEoverridecommandlockouts                              % This command is only needed if 
\usepackage{graphicx} % for pdf, bitmapped graphics files
\usepackage{epsfig} % for postscript graphics files
\usepackage{mathptmx} % assumes new font selection scheme installed
\usepackage{times} % assumes new font selection scheme installed
\usepackage{amsmath} % assumes amsmath package installed
\usepackage{amssymb}  % assumes amsmath package installed
\usepackage{textgreek}
\usepackage{hyperref}
\binoppenalty=\maxdimen
\relpenalty=\maxdimen
\title{\LARGE \bf
Dynamic Visualization of Gyral and Sulcal Stereoelectroencephalographic contacts in Humans 
}

\author{Markus Adamek$^{1,2}$, Alexander P Rockhill$^{4}$, Peter Brunner$^{1,2}$ and Dora Hermes$^{3}$% <-this % stops a space
\thanks{$^{1}$Department of Neurosurgery, Washington University in Saint Louis, MO, USA}%
\thanks{$^{2}$National Center for Adaptive Neurotechnologies, Albany, NY, USA}%
\thanks{$^{3}$Department of Physiology \& Biomedical Engineering, Mayo Clinic Rochester, MN, USA}%
\thanks{$^{4}$University of Oregon, Department of Human Physiology, Eugene, OR, USA}%
}

\begin{document}

\maketitle
\thispagestyle{empty}
\pagestyle{empty}

%%%%%%%%%%%%%%%%%%%%%%%%%%%%%%%%%%%%%%%%%%%%%%%%%%%%%%%%%%%%%%%%%%%%
\begin{abstract}

Stereoelectroencephalography (SEEG) is a neurosurgical method to survey electrophysiological activity within the brain to treat disorders such as Epilepsy. In this stereotactic approach, leads are implanted through straight trajectories to survey both cortical and sub-cortical activity.

Visualizing the recorded locations covering sulcal and gyral activity while staying true to the cortical architecture is challenging due to the folded, three-dimensional nature of the human cortex. 

To overcome this challenge, we developed a novel visualization concept, allowing investigators to dynamically morph between the subjects' cortical reconstruction and an inflated cortex representation. This inflated view, in which gyri and sulci are viewed on a smooth surface, allows better visualization of electrodes buried within the sulcus while staying true to the underlying cortical architecture. 
\newline

\indent \textit{Clinical relevance}—
These visualization techniques might also help guide clinical decision-making when defining seizure onset zones or resections for patients undergoing SEEG monitoring for intractable epilepsy.

\end{abstract}

%%%%%%%%%%%%%%%%%%%%%%%%%%%%%%%%%%%%%%%%%%%%%%%%%%%%%%%%%%%%%%%%%%%%%%%%%%%%%%%%
\section{INTRODUCTION}
Behavior emerges from complex interactions between functionally specialized regions in the brain. Uncovering these regional specializations and the complex networks required to produce behavior have been of scientific and clinical interest for decades. For example, electrode grids placed directly on the surface of the human cortex (Electrocardiography; ECoG) are used to determine the optimal treatment for patients with epilepsy while creating a unique opportunity to investigate human visual, motor, and auditory networks \cite{Crone2006, Miller2007, Swift2018}.

\begin{figure}[h]
\centering
  \vspace{5pt} % adding some V space because of margins
  \includegraphics[width=0.40\textwidth]{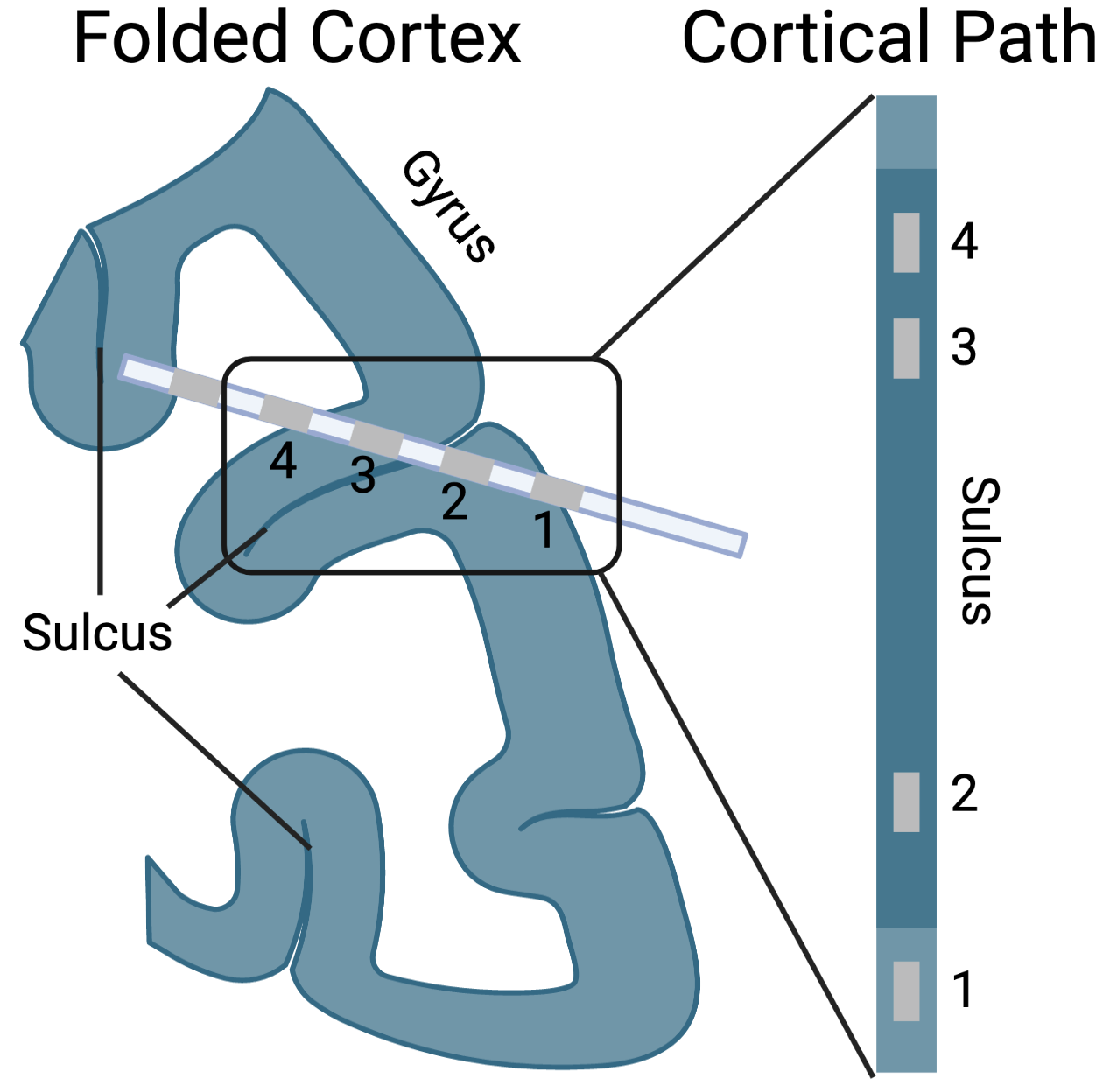}
  \caption{\textbf{Schematic representation electrode location distance issue in Euclidean space.} The left image shows a hypothetical SEEG trajectory passing through the folded cortex. SEEG recording electrodes are equally spaced along the trajectory. However, if the cortical grey matter is stretched out (as it would be in an inflated representation), the distances between the recording locations change. For example, the distance between electrodes 2 and 3 increases since the sulcus separates them. Note that the trajectory crosses a sulcus to demonstrate how sulci and gryi are represented on the inflated brain, not as a realistic trajectory.}
  \label{fig:figure1}
\end{figure}

In recent years, Stereoelectroencephalography (SEEG) has started to replace ECoG in the United States. In the SEEG approach, leads are implanted through straight trajectories, surveying both cortical and sub-cortical areas. Therefore, the electrodes placed along the trajectory will pass through the highly folded human cortex  (\autoref{fig:figure1}), opening up the possibility of investigating the functional differences between electrodes recording sulcal and gyral activity with high fidelity. This paradigm shift requires us to rethink how we localize, visualize and analyze the activity recorded through this approach, creating a new set of challenges. 

First, visualizing the location of electrodes within the sulcus is difficult. Investigators and clinicians have to look either at (1) 2-D slice stacks or (2) semi-transparent 3D models, which are both impractical for different reasons. While looking at electrode locations on the anatomical 2-D slices provides an accurate anatomical description for a specific contact, it is impractical to investigate widespread network activity, as there is no conceivable way to visualize all electrodes simultaneously. This issue can be somewhat addressed by viewing the electrode locations on a semi-transparent reconstructed 3D model. However, two electrodes that appear close to each other in the semi-transparent brain might be considered far apart if the folded nature of the cortex is taken into account. (\autoref{fig:figure1}).  

Secondly, the recorded activity needs to be viewed in a space that allows accurate association between the recorded data and the anatomical and functional architecture of the brain. There is mounting evidence that sulcal and gyral  cortical grey matter are functionally distinct \cite{deng_functional_2014,jiang_fundamental_2021}. Understanding the temporal spread of activation across gyri and sulci, therefore, depends on an accurate representation of the folded cortex. \cite{Crowther2019, Moheimanian2021, Coon2016}. 

%placeholder: https://academic.oup.com/psyrad/article/1/1/23/6187507
To address these issues, we developed a novel approach for localizing, visualizing, and analyzing SEEG data. In this approach, we morph the 3D surface, and associated electrode locations from its gyrated 3D model to an inflated model \cite{dale_cortical_1999,fischl_cortical_1999} through models created by Freesurfer. In the inflated model, the cortical surface is smooth, allowing visualization of sulci and gyri while maintaining their topological structure. %This becomes especially important considering the functional differences moving from the sulcus to the gyrus to the sulcus.

% https://doi.org/10.1006/nimg.1998.0396
This approach allows visualization of electrode locations in the original anatomically accurate 3D space with the ability to slowly morph the model into a view representing the position of electrodes on the inflated model, enabling a better view from a more functional perspective. 

%%give some anatomical examples where this is important? auditory cortex for example!
   
\section{METHODS}

\subsection*{Surface Reconstruction}
The surface and the inflated surface were reconstructed from the subjects' T1-weighted MRI using Freesurfer. In short, the intensity of the T1-weighted MRI is normalized, and non-cerebral voxels are removed. The resulting image is then processed further to remove subcortical components, and finally, a surface mesh of the cortex is created. Next, Freesurfer calculates the inflated model by minimizing the number of folds on the surface. The resulting surface meshes have a 1:1 vertex identity, allowing a surface vertex to be identified in the inflated surface.

\subsection*{Implementation}

The visualization is available in MATLAB and Python, using Freesurfers' reconstructed surface and inflated model. Both implementations are freely available on GitHub. The Matlab version was implemented as a standalone package\cite{Adamek_Dynamic_Visualization_of_2023} and subsequently integrated into the Versatile Electrode Localization Framework (VERA) \cite{Adamek_VERA_-_A_2022}. The implementation can also be used independently of VERA, allowing integration into existing workflows. In addition to the 3D model morphing, points (through scatter3) or text can also be morphed. The implementation enables linking these different visualizations to ensure that morphing one object results in correct morphs of all linked objects. Furthermore, the code is easily extendable to create additionally linked visualizations.

The Python version was implemented as part of the  MNE-Python package \cite{GramfortEtAl2013a}. The MNE-Python implementation can be seen here (\href{https://mne.tools/dev/auto_tutorials/clinical/20_seeg.html}{https://mne.tools/dev/auto\_tutorials/clinical/20\_seeg.html}).

\subsection*{Electrode Locations}
Electrode locations were reconstructed from a post-op CT using VERA. The CT was first co-registered with the T1-weighted MRI used for surface reconstruction. Next, we use VERA's fully-automated segmentation algorithm to determine the electrode locations from the CT, followed by manual correction.

\subsection*{Algorithm}

We morph between the surface and inflated model using a morphing parameter $0 \leq \sigma \leq 1 $. Assuming the $i^{th}$ vertex $p_{ci}$ of the cortical surface corresponds to the vertex $i^{th}$ vertex $p_{ii}$ on the inflated model, the morphed vertex location $p_{mi}$ is calculated as \begin{align*}
    p_{mi}=(1-\sigma) p_{ci}+\sigma p_{ii}
\end{align*}
Next, we determined which electrodes are within the cortical surface matter versus those in subcortical structures or white matter. This can be achieved through multiple avenues. The simplest method is to define a distance threshold $d$ so that only electrodes $e_j$ remain, which satisfies the condition 
\begin{align*}
d_{j}=\min\limits_{\forall i} \lVert e_j - p_{ci} \rVert < d 
\end{align*} 
In our case, the distance threshold was set to $d=4mm$. Therefore, any electrode $e_j$ close enough to the cortical surface will be visualized. However, cortical thickness varies across the cortex as well as between subjects. An alternative method is to determine electrode locations through volume-derived labels. 

Finally, for each electrode location $e_j$, we determine the closest cortical vertex $p_{ci_j}$ and its associated inflated vertex location $p_{ii_j}$. The inflated vertex location $p_{ii_j}$ is then used to determine the morphed electrode location $e_{mj}$. \begin{align*}
	   i_j=\arg\min_{\forall i} (p_{ci} - e_j)\\
          e_{mj}=(1-\sigma)e_{j}+\sigma p_{ii_j}
\end{align*}

\section{RESULTS}

\begin{figure*}[ht]
  \centering
  \vspace{5pt}
  \includegraphics[width=0.90\textwidth]{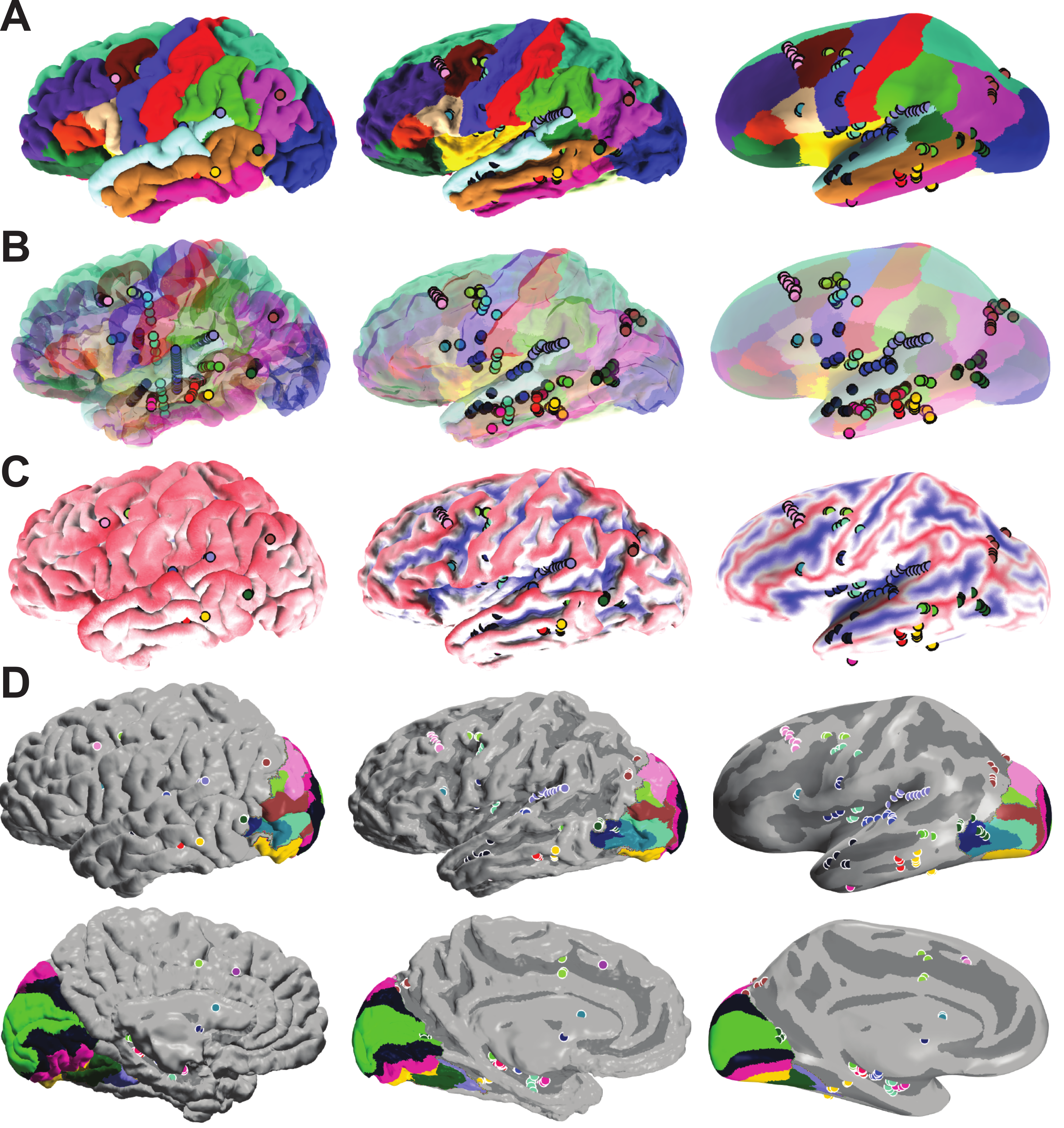}
  \caption{\textbf{Example of an SEEG subject being morphed from the 3D Surface to the inflated model.}
  Three different visualizations of the same subject at different morphing steps. 
  \textbf{(A)} 
  Reconstructed and labeled surfaces, including implanted electrode locations. As can be seen here, without inflation, almost non of the implanted electrodes can be visualized as the gyrated surface hides them. As we inflate the surface, we can visualize electrodes recording sulcal activity.
  \textbf{(B)} Semi-transparent reconstructed cortical surface including implanted electrode trajectories. While a semi-transparent surface allows visualization of deeper electrodes, identifying the anatomical position of an implanted electrode is challenging.
  \textbf({C}) In this view, the gyri are labeled in red, and the sulci are labeled in blue. As expected, in the uninflated view, no sulci are visible.
  \textbf{(D)} Visualization of visual fields from a lateral (top) and medial (bottom) view. Locations of electrodes within PHC1 (dark blue, top) and V1v (neon green, bottom) can only be identified after inflating the cortex.
  }
  \label{fig:figure2}
\end{figure*}

To illustrate the issues solved by our implementation, we present four scenarios. The first scenario is the classical view of the cortex on a non-transparent cortical reconstruction. Without transparency, only a few of the 138 contacts (The patient was implanted with 231 electrodes, and 138 passed our $4mm$ distance criteria) are visible in \autoref{fig:figure2}~A. 
However, after morphing the surface and electrode locations, we can determine the anatomical identity of all cortical electrodes (\autoref{fig:figure2}~A,~\textsigma=1). 

In the second scenario, we use a semi-transparent surface to illustrate the issue of three-dimensional representations. While in this view (\autoref{fig:figure2}~B), all electrodes are visible, without morphing, it is hard to determine their anatomical identity. 

Representing the same subject but color-code sulci and gyri helps illustrate the issue of identifying recording locations buried within the sulci. As expected, sulci cannot be visualized without morphing, and electrodes recording activity within these sulci cannot be observed (\autoref{fig:figure2}~C).

Lastly, we overlapped the sulcus map with visual field areas identified via the neuropythy package \cite{10.7554/eLife.40224}.  Without the inflated model, identifying the electrodes which are probable to record activity from V1v (neon green) and PHC1 (dark blue) would not be possible (\autoref{fig:figure2}~D).

\section{CONCLUSIONS}

Here we present a novel method to visualize the relationship between electrode locations and their anatomical recording location for patients implanted with SEEG electrodes. 

The shift from ECoG to SEEG has opened up new clinical and research opportunities but also requires us to rethink the methods we use to visualize and analyze said activity. These opportunities span a wide range of active research areas, all of which will benefit from the additional information provided by SEEG.

One of these active research areas investigates the temporal dynamics of auditory processing. While some research suggests a caudal to rostral spread of cortical activity during auditory information processing, others refute this hypothesis \cite{Hamilton2018, Nourski2014, Camalier2012}. However, evidence for this hypothesis of human auditory processing in humans was primarily investigated via ECoG grids, which do not record sulcal activity. SEEG activity, combined with the presented visualization method, might elucidate current debates by taking to account the complex underlying cytoarchitectural differences between sulci and gyri in the auditory cortex \cite{Zachlod2020}.

Similarly, visual processing is distributed smoothly across specific gyral and sulcal locations \cite{Swisher2007, Wandell2007, Wandell2011}. Therefore, knowledge of the location of a recording electrode is of paramount importance. 

And lastly, the proposed visualization could help understand the underlying mechanism of traveling waves. These spatially organized electrophysiological patterns have been identified as essential patterns, organizing information flow across the cortex\cite{Ermentrout2001, Muller2018, Das2022, Bhattacharya2022}. However, the idea of traveling waves is based on a flat cortical surface on which the wave propagates, necessitating an inflated brain model to examine traveling waves truthfully.

\addtolength{\textheight}{0cm}   % This command serves to balance the column lengths

%%%%%%%%%%%%%%%%%%%%%%%%%%%%%%%%%%%%%%%%%%%%%%%%%%%%%%%%%%%%%%%%%%%%%%%%%%%%%%%%

%%%%%%%%%%%%%%%%%%%%%%%%%%%%%%%%%%%%%%%%%%%%%%%%%%%%%%%%%%%%%%%%%%%%%%%%%%%%%%%%

%%%%%%%%%%%%%%%%%%%%%%%%%%%%%%%%%%%%%%%%%%%%%%%%%%%%%%%%%%%%%%%%%%%%%%%%%%%%%%%%

\section*{ACKNOWLEDGMENT}

Research reported in this publication was supported by the National Institute Of Mental Health of the National Institutes of Health under Award Numbers R01-MH122258 (DH, PB), R01-EB026439 (PB), P41-EB018783 (PB), U24-NS109103 (PB), U01-NS108916 (PB) and U01-NS128612 (PB, DH). The content is solely the responsibility of the authors and does not necessarily represent the official views of the National Institutes of Health.
\autoref{fig:figure1} was created with BioRender.com.

%%%%%%%%%%%%%%%%%%%%%%%%%%%%%%%%%%%%%%%%%%%%%%%%%%%%%%%%%%%%%%%%%%%%%%%%%%%%%%%%

%\renewcommand{\bibsection}{} % remove the references title from bibliography
\bibliographystyle{IEEEtran}
\bibliography{bibliography}

\end{document}